\documentstyle[pra,aps,multicol,epsfig]{revtex}

\input epsf.tex

\newcommand{\be}{\begin{equation}}
\newcommand{\ee}{\end{equation}}
\newcommand{\ba}{\begin{eqnarray}}
\newcommand{\ea}{\end{eqnarray}}
\newcommand{\ban}{\begin{eqnarray*}}
\newcommand{\ean}{\end{eqnarray*}}

\newcommand{\sandwich}[3]{\mbox{$ \langle #1 | #2 | #3 \rangle $}}
\newcommand{\ket}[1]{\mbox{$ | #1 \rangle $}}
\newcommand{\bra}[1]{\mbox{$ \langle #1 | $}}

\newcommand{\si}{\sigma}
\newcommand{\demi}{\frac{1}{2}}

\newcommand{\one}{\leavevmode\hbox{\small1\normalsize\kern-.33em1}}

\newcommand{\moy}[1]{\langle #1 \rangle}

\begin{document}

\title{Bell-type inequalities to detect true $n$-body non-separability}
\author{Daniel Collins$^{1,2}$, Nicolas Gisin$^3$,
Sandu Popescu$^{1,2}$, David Roberts$^1$, Valerio Scarani$^3$ }
\address{
$^1$ H.H. Wills Physics Laboratory, University of Bristol, Tyndall
Avenue, Bristol BS8 1TL, UK\\
$^2$ BRIMS, Hewlett-Packard Laboratories, Stoke Gifford, Bristol BS12 6QZ, UK\\
$^3$ Group of Applied Physics, University of Geneva, 20, rue de
l'Ecole-de-M\'edecine, CH-1211 Geneva 4, Switzerland}
\date{14 January 2002}
\maketitle

\begin{abstract}
We analyze the structure of correlations among more than two
quantum systems. We introduce a classification of correlations
based on the concept of non-separability, which is different {\em
a priori} from the concept of entanglement. Generalizing a result
of Svetlichny [Phys. Rev. D {\bf 35} (1987) 3066] on 3-particle
correlations, we find an inequality for $n$-particle correlations
that holds under the most general separability condition and that
is violated by some quantum-mechanical states.
\end{abstract}

\begin{multicols}{2}

Quantum mechanics (QM) predicts remarkable correlations between
the outcomes of measurements on sub-systems (particles) of a
composed system. This prediction is consequence of the linearity
of QM, that allows to build superposition states that cannot be
written as products of states of each sub-system. Such states are
called {\em entangled}. Entanglement is at the heart of some of
the most puzzling features of QM: the Einstein-Podolski-Rosen
argument, the measurement problem, the paradox of Schr\"{o}dinger
cat... In the last decade, it has been noticed that entanglement
is also a {\em resource} that allows to perform tasks that would
be classically impossible. This new, more effective standpoint
caused a renewed interest in the study of quantum correlations
\cite{machines}.

The correlations among more than two quantum systems have been the
object of several recent studies, also motivated by the fact that
experiments aimed to check such correlations are becoming
feasible. Usually, the structure of correlation has been
classified according to {\em entanglement}. In this Letter, we
propose a complementary classification, in terms of {\em
non-separability}. Before entering the technicalities, it is
useful to explain why the concepts of entanglement and of
non-separability are {\em a priori} different concepts. We do this
on the simplest case, that of correlations between three
particles.

The {\em classification through entanglement} presupposes that the
system of three particles admits a quantum-mechanical description.
Thus, any state of the system is described by a density matrix
$\rho$. To classify a given $\rho$ in terms of entanglement, one
must consider all possible decompositions of the state as a
mixture of pure states $\rho=\sum_ip_i\,\ket{\Psi_i}\bra{\Psi_i}$.
Then (i) If there exist a decomposition for which {\em all}
$\ket{\Psi_i}$ are product states
$\ket{\psi_i^1}\ket{\psi_i^2}\ket{\psi_i^3}$, then $\rho$ is not
entangled at all, that is, it can be prepared by acting on each
subsystem separately. For this situation, we use the acronym {\em
1/1/1QM}. (ii) If all $\ket{\Psi_i}$ can be written as either
$\ket{\psi_i^{12}}\ket{\psi_i^3}$ or
$\ket{\psi_i^{13}}\ket{\psi_i^2}$ or
$\ket{\psi_i^{23}}\ket{\psi_i^{1}}$, and at least one of the
$\ket{\psi_i^{jk}}$ is not a product state, then $\rho$ is
entangled, but there is no true three-particle entanglement. We
shall say that $\rho$ exhibits two-particle entanglement, and use
the acronym {\em 2/1QM} to refer to it. (iii) Finally, if for any
decomposition there is at least one $\ket{\Psi_i}$ that shows
three-particle entanglement, then to prepare $\rho$ one must act
on the three subsystems: $\rho$ exhibits true three-particle
entanglement (acronym {\em 3QM}).

It is difficult to establish to which class a given $\rho$
belongs, because in principle one should write down {\em all} the
possible decompositions of $\rho$ onto pure states. In fact, to
date no general criterion is known. However, we know a sufficient
criterion: there exists an operator ${\cal{M}}_3$ such that: (a) if
$\mbox{Tr}(\rho{\cal{M}}_3)>1$, then certainly $\rho$ is
entangled; (b) if $\mbox{Tr}(\rho{\cal{M}}_3)>\sqrt{2}$, then
certainly $\rho$ exhibits true three-particle entanglement. The
operator ${\cal{M}}_3$ is the Bell operator that defines the
so-called Mermin inequality \cite{mermin}; we shall come back to
it later.

The {\em classification through non-separability} (or
non-locality) does not presuppose that the system of three
particles admits a quantum-mechanical description. Rather, we have
the following cases: 

(i) Each particle separately carries a script
$\lambda$, which determines the outcome of each possible
measurement. When the experiment is repeated, the script can be
different; the script $\lambda$ occurring with probability
$\rho(\lambda)$. This is the so-called {\em local variables} ({\em
lv}) model, which we will also denote as {\em 1/1/1S}. More
exactly, in the {\em lv} model the joint probabilities $P(a_1,
a_2,a_3)$ that an arbitrary experiment $A_1$ performed on particle
$1$ yields the result $a_1$, while the measurement $A_2$ performed on
particle $2$ yields $a_2$ and the measurement $A_3$ on particle $3$ yields
$a_3$ is given by:
\begin{equation}
P(a_1,a_2,a_3) = \int d\lambda \rho(\lambda) P_1(a_1|
\lambda)P_2(a_2| \lambda)P_3(a_3| \lambda)\label{local} 
\end{equation} 
where $P_1(a_1| \lambda)$ is the probability that
when particle 1 carrying the script $\lambda$ is subjected to a
measurement of $A_1$ it yields the result $a_1$, and similarly for $P_2$
and $P_3$. Note that {\em lv} is more general than {\em 1/1/1QM}.
For instance, in {\em lv} one can build a state such that the
measurement of the spin of particle 1 along both directions
$\hat{z}$ and $\hat{x}$ gives +1 with certainty, while such a
state does not exist in QM.

(ii) The intermediate case, first considered by
Svetlichny\cite{svet}, is an {\em hybrid local - nonlocal model}:
for each triple of particles, we allow arbitrary (i.e. nonlocal)
correlation between two of the three particles, but only local
correlations between these two particles and the third one; which
pair of particles is nonlocally correlated may be different in
each repetition of the experiment.  If we define $p_{i,j}$ to be 
the probability that particles i and j are nonlocally correlated,
then in this model
\begin{eqnarray}
P_{1,2,3}(a_1,a_2,a_3)  &=& \sum_{k=1}^3 \, p_{i,j}\,\int d\lambda
\big[\rho_{i,j}(\lambda) \nonumber \\
&& P_{i,j}(a_i,a_j|\lambda)P_k(A_k=a_k| \lambda)\big],\label{hybrid}
\end{eqnarray}
where $\{i,j,k\}$ is an even permutation of $\{1,2,3\}$.
We refer to this situation by the acronym {\em 2/1S}.
Note again that {\em 2/1S} is more general than {\em 2/1QM}, since
we don't require that the two correlated particles are correlated
according to QM.

(iii) The last situation ({\em 3S}) is the one without
constraints: we allow all the three particles to share an
arbitrary correlation.

It is not evident {\em a priori} whether three-particle
entanglement {\em 3QM} is stronger, equivalent or weaker than {\em
2/1S}. The proof that {\em 3QM} is actually {\em stronger} than
{\em 2/1S} was given some years ago by Svetlichny \cite{svet}, who
found an inequality for three particles that holds for {\em 2/1S}
and is violated by QM. In this Letter, we are going to exhibit a
generalized Svetlichny inequality for an arbitrary number of
particles $n$, that is, an inequality that allows to discriminate
$n$-particle entanglement {\em nQM} from any hybrid model {\em
k/(n-k)S}.

The plan of the paper is as follows. First, we introduce the
family of the Mermin-Klyshko (MK) inequalities \cite{mermin,mk},
that will be the main tool for this study. With this tool, we
re-derive Svetlichny's inequality for three particles and compare
it to Mermin's. We move then to the case of four particles, and
show that the MK inequality plays the role of generalized
Svetlichny inequality. Finally, we generalize our results for an
arbitrary number of particles $n$.\\

We consider from now onwards an experimental situation in which
{\em two dichotomic measurements} $A_j$ and $A_j'$ can be
performed on each particle $j=1,...,n$. The outcomes of these
measurements are written $a_j$ and $a'_j$, and can take the values
$\pm 1$.  Letting $M_1=a_1$, we can define recursively the
{\em MK polynomials} as \ba M_n&=&\demi\,M_{n-1}\,(a_n+a_n')\,+\,
\demi\,M'_{n-1}\,(a_n-a_n')\,, \label{mkpol}\ea where $M_k'$ is
obtained from $M_k$ by exchanging all the primed and non-primed
$a$'s. In particular, we have \ba
M_2&=&\demi\big(a_1a_2+a_1'a_2+a_1a_2'-a_1'a_2'\big)\,,
\label{chshpol}\\
M_3&=&\demi\big(a_1a_2a_3'+a_1a_2'a_3+a_1'a_2a_3-a_1'a_2'a_3'\big)\,.\label{mermpol}\ea
The recursive relation (\ref{mkpol}) gives, for all $1\leq k\leq
n-1$: \ba M_n&=& \demi\,M_{n-k}\,(M_k+M_k')\,+\,
\demi\,M'_{n-k}\,(M_k-M_k')\,.\label{mkpol2}\ea 
We shall interpret these ploynomials as sums of expectation values, 
eg. we shall interpret $M_2$ as 
\begin{equation}
\frac{1}{2}\left(E(A_1A_2)+E(A_1'A_2)+E(A_1A_2')-E(A_1'A_2')\right),
\end{equation}
where $E(A_1A_2)$ is the expectation value of the product 
$A_1A_2$ when $A_1$ and $A_2$ are measured (note that $A_1$ and
$A_1'$ cannot be measured at the same time).  We call quantities
such as $E(A_1A_2A_3)$ correlation coefficients.  We shall look at
the values of these polynomials under QM and hybrid local/non-local
variable models, and show that they give generalised Bell inequalities.

We shall first look at hybrid local/non-local variable models.  
For technical simplicity, throughout this paper we consider only
{\em deterministic} versions of the hybrid variable models,
which means that the script $\lambda$ in eq. (\ref{local}) and
(\ref{hybrid}) {\em completely} determines the outcome of the
measurements (i.e the probabilities $P_1(X=x|\lambda)$ and similar
are either zero or one). It is known that any non-deterministic
local variable model can be made deterministic by adding
additional variables\cite{note2}.  In addition, we can also
use the script $\lambda$ to determine which particles are allowed
to communicate non-locally: eg. for 3 parties the probabilites are
now given simply by
\begin{equation}
P(a_1,a_2,a_3)  = \int d\lambda \rho(\lambda) P(a_1,a_2,a_3|\lambda), 
\end{equation}
where for each $\lambda$ the probabilities must factorize as 
some 2/1 grouping (though not necessarily the same for different   
$\lambda$).  Now, for any $\lambda$, the outcomes of all products
$A_1A_2A_3$ etc. are fixed, and so we can define the fixed 
quantity $M_n^{\lambda}$.  The value of $M_n$ is just
the probabilistic average over $\lambda$ of $M_n^{\lambda}$.  Thus
if we can put a bound upon all possible $M_n^{\lambda}$ 
then we have a bound upon $M_n$.  For example, it can be shown that for any
{\em lv} model, $M_n\leq 1$.  This can be easily seen from
(\ref{mkpol}) using a recursive argument, noting that for any
script of local variables it holds that either $a_n=a_n'$ or
$a_n=-a_n'$. In particular, $M_2\leq 1$ for {\em lv} is the
Clauser-Horne-Shimony-Holt inequality for two particles
\cite{chshpaper}. On the opposite side, if we consider the model
without constraints {\em nS}, then $M_n$ can reach the so-called
{\em algebraic limit} $M_n^{alg}$, achieved by setting at +1
(resp. -1) all the correlations coefficients that appear in $M_n$
with a positive (resp. negative) sign. So for example
$M_2^{alg}=M_3^{alg}=2$.

Turning to QM: since we consider dichotomic measurements, we can
restrict to the case of two-dimensional systems (qubits)\cite{werner}. 
In this
case, the observable that describes the measurement $A_j$ can be
written as $\vec{a_j}\cdot\vec{\si}\equiv \si_{a_j}$, with
$\vec{a_j}$ a unit vector and $\vec{\si}$ the Pauli matrices. The
equivalent of $M_n$ is the expectation value of the operator
${\cal{M}}_n$ obtained by replacing all $a$'s by the corresponding
$\si_a$. It is thus known that QM violates the inequality
$\mbox{Tr}(\rho\,{\cal{M}}_n)\leq 1$. More precisely, it is known
\cite{mk} that: (I) The maximal value achievable by QM is
$\mbox{Tr}(\rho\,{\cal{M}}_n)= 2^{\frac{n-1}{2}}$, reached by the
generalized Greenberger-Horne-Zeilinger (GHZ) states
$\frac{1}{\sqrt{2}}\big(\ket{0...0}+\ket{1...1}\big)$; (II) If
$\rho$ exhibits $m$-particle entanglement, with $1\leq m\leq n$,
then $\mbox{Tr}(\rho\,{\cal{M}}_n)\leq 2^{\frac{m-1}{2}}$
\cite{note1}. In other words, if we have a state of $n$ qubits
$\rho$ such that $\mbox{Tr}(\rho\,{\cal{M}}_n)>
2^{\frac{m-1}{2}}$, we know that this state exhibits at least
$(m+1)$-particle entanglement. This means that the MK-polynomials
allow a classification of correlations according to entanglement.
But do they allow also the classification according to
non-separability? The answer to this question is: yes for $n$
even, no for $n$ odd. As announced, we demonstrate this statement
first for $n=3$, then for $n=4$, and finally for all $n$.\\

{\em Three particles.} Let's take the Mermin polynomial $M_3$
given in (\ref{mermpol}). We have already discussed the following
bounds: $M_3^{lv}=1$, $M_3^{2/1QM}=\sqrt{2}$,
$M_3^{3QM}=M_3^{alg}=2$. We lack the bound for {\em 2/1S}. This is
easily calculated: consider a script in which particles 1 and 2
are correlated in the most general way, and particle 3 is
uncorrelated with the others. Then we use (\ref{mkpol}), that
reads $M_3=\demi\,M_{2}\,(a_3+a_3')\,+\,
\demi\,M'_{2}\,(a_3-a_3')$. For any particular script, as we said
above, $a_3$ can only be equal to $\pm a_3'$. Without loss of
generality, we choose $a_3=a_3'=1$, whence $M_3^{2/1S}=\max M_2$.
Since particles 1 and 2 can have the highest correlation, $\max
M_2=M_2^{alg}$ here. In conclusion, $M_3^{2/1S}=2$. Thus, for
Mermin's polynomial \be
M_3^{2/1S}\,=\,M_3^{3QM}\,=\,M_3^{alg}\,=\,2\,: \ee the Mermin
polynomial does not discriminate between the deterministic
variable models {\em 2/1S} and {\em 3S}, and the
quantum-mechanical correlation due to three-particle entanglement.
The deep reason for this behavior lies in the fact that $M_3$ has
only four terms: the correlations $a_1a_2a_3$, $a_1'a_2'a_3$,
$a_1'a_2a_3'$ and $a_1a_2'a_3'$ do not appear in $M_3$ given in
(\ref{mermpol}). But these correlations are those that appear in
$M_3'$; thus we are lead to check the properties of the polynomial
\ba S_3&=&\demi(M_3+M_3')\,=\,\demi(M_2 a_3'+M_2' a_3)\,. \ea For
both {\em lv} and {2/1S}, the calculation goes as follows: we
choose $a_3=a_3'=1$, and we are left with
$S_3^{...}=\demi\,\max(M_2+M_2')$. But $M_2+M_2'=a_1a_2'+a_1'a_2$,
which can take the value of 2 in both {\em lv} and {2/1S}.
Therefore $S_3^{lv}=S_3^{2/1S}=1$, and this implies immediately
$S_3^{2/1QM}=1$ since {\em 2/1QM} is more general than {\em lv}
and is a particular case of {\em 2/1S}. The algebraic maximum is
obviously $S_3^{alg}=2$. We have to find $S_3^{3QM}$. As above, we
define an operator ${\cal{S}}_3$ by replacing the $a$'s in the
polynomial $S_3$ with Pauli matrices. On the one hand, we have \ba
\mbox{Tr}(\rho{\cal{S}}_3)&=&\demi\big[\mbox{Tr}(\rho\,{\cal{M}}_2\si_{a_3'})
+ \mbox{Tr}(\rho\,{\cal{M}}_2'\si_{a_3}) \big]\,\leq \,\sqrt{2}
\ea since by Cirel'son theorem \cite{cirel} each term of the sum
is bounded by $\sqrt{2}$. On the other hand, we know \cite{sca1}
that the eigenvector associated to the maximal eigenvalue for such
an operator is the GHZ state $\frac{1}{\sqrt{2}}\big(
\ket{000}+\ket{111}\big)$. For some settings \cite{settings}, we
have $\sandwich{GHZ}{{\cal{S}}_3}{GHZ}=\sqrt{2}$: the bound can be
reached, that is, $S_3^{3QM}=\sqrt{2}$.  Thus the GHZ state 
generates genuine $3$-party non-separability (non-locality).  We 
note that, in fact, $S_3$ is one of 
Svetlichny's two inequalities [the second inequality is
equivalent, and is associated to $\demi(M_3-M_3')$].

The results for Mermin's and Svetlichny's inequalities for three
particles are summarized in Table \ref{tablefin}. We see that
combining Mermin's and Svetlichny's inequalities one can
discriminate between the five models for correlations that we
consider in this paper. This concludes our study of the case of
three particles.\\

{\em Four particles.} As above, we begin by considering the MK
polynomial $M_4$. Like $M_2$, and unlike $M_3$, the polynomial
$M_4$ is a linear combination of the correlation coefficients of
{\em all} measurements. From the general properties of the MK
inequalities \cite{mk}, the following bounds are known:
$M_4^{lv}=1$, $M_4^{1/1/2QM}=M_4^{2/2QM}=\sqrt{2}$,
$M_4^{3/1QM}=2$, $M_4^{4QM}=2\sqrt{2}$. The algebraic limit is
$M_4^{alg}=4$ (sixteen terms in the sum, and a factor
$\frac{1}{4}$ in front of all).

Now we have to provide the bounds for {\em 1/1/2S}, {\em 2/2S} and
{\em 3/1S}. This last one can be calculated in the same way as
above: using (\ref{mkpol}), we have
$M_4=\demi\,M_{3}\,(a_4+a_4')\,+\, \demi\,M'_{3}\,(a_4-a_4')$; we
set $a_4=a_4'=1$, and since we allow the most general correlation
between the first three particles we have $\max M_3=M_3^{alg}=2$.
Therefore $M_4^{3/1S}=2$.

One must be more careful in the calculation of {\em 1/1/2S} and
{\em 2/2S}. This goes as follows: using (\ref{mkpol2}), we have
$M_4= \demi\,M_{1,2}\,(M_{3,4}+M_{3,4}')\,+\,
\demi\,M'_{1,2}\,(M_{3,4}-M_{3,4}')$, where to avoid confusions we
wrote $M_{i,j}$ instead of $M_2$, with $i$ and $j$ the labels of
the particles. Now, $M_{3,4}+M_{3,4}'=a_3a_4'+a_3'a_4$, and
$M_{3,4}-M_{3,4}'=a_3a_4-a_3'a_4'$. So if we allow the most
general correlation between particles 3 and 4, these two
quantities are independent and can both reach their algebraic
limit, which is 2. Consequently for both {\em 1/1/2S} and {\em
2/2S} we obtain $M_4^{...}=\max(M_{1,2}+M'_{1,2})$, which is again
2 in both cases. So finally \ba
M_4^{1/1/2S}=M_4^{2/2S}=M_4^{3/1S}=2&<& M_4^{4QM}=2\sqrt{2}\,: \ea
for four particles, the MK polynomial $M_4$ detects both
four-particle entanglement (this was known) and four-particle
non-separability, and is therefore
the natural generalization of Svetlichny's inequality.\\

{\em Arbitrary number of particles.} For a given number of
particles $n$, we discuss only the maximal value allowed by QM,
that is the case {\em nQM}, against any possible partition in {\em
two} subsets of $k$ and $n-k$ particles respectively, with $1\leq
k\leq n-1$, that is the case {\em k/(n-k)S}. Partitions in a
bigger number of smaller subsets are clearly special cases of
these bilateral partitions. We are going to prove the following

{\bf Proposition:} {\em Define the generalized Svetlichny
polynomial $S_n$ as \ba S_n&=& \left\{ \begin{array}{lcl} M_n
&,&\mbox{$n$ even}\\ \demi(M_n+M_n') &,&\mbox{$n$ odd}
\end{array}
\right.\,. \ea Then all the correlations {\em k/(n-k)S} give the
same bound $S_n^{k}$, and the bound that can be reached by QM is
higher by a factor $\sqrt{2}$:} \ba
S_n^{nQM}&=&\sqrt{2}\,S_n^{k}\,. \ea

The tools for the demonstration are the generalization to all the
MK polynomials of the properties of $M_2$ and $M_3$ that we used
above, namely: (I) the algebraic limit of $M_k$ is:
$M_k^{alg}=2^{\frac{k}{2}}=M_k^{kQM}\sqrt{2}$ for $k$ even, and
$M_k^{alg}=2^{\frac{k-1}{2}}=M_k^{kQM}$ for $k$ odd. (IIa) For $k$
even, $M_k$ and $M_k'$ are different combinations of all the
correlation coefficients; $M_k+M_k'$ and $M_k-M_k'$ contain each
one half of the correlation coefficients, and the algebraic limit
for both is $M_k^{alg}$. (IIb) For $k$ odd, $M_k$ and $M_k'$
contain each one half of the correlation coefficients. These
properties are not usually given much stress, but can indeed be
found in \cite{mk}, or easily verified by direct inspection.

Let's first prove the Proposition for $n$ even. In this case, the
QM bound is known to be $S_n^{nQM}=2^{\frac{n-1}{2}}$. As in the
case of four particles, to calculate $S_n^k$ we must distinguish
two cases:
\begin{itemize}
\item For $k$ and $n-k$ even: in (\ref{mkpol2}), both
$M_k+M_k'$ and $M_k-M_k'$ can be maximized independently because
of property (IIa) above; therefore, we replace them by
$M_k^{alg}$. We are left with $S_n^k=\demi
M_k^{alg}\,\max(M_{n-k}+M'_{n-k})$, and this maximum is again
$M_{n-k}^{alg}$. So finally $S_n^k=\demi M_k^{alg} M_{n-k}^{alg} =
2^{\frac{n-2}{2}}$.
\item For $k$ and $n-k$ odd: in (\ref{mkpol2}), $M_{n-k}$ and
$M_{n-k}'$ can be optimized independently because of (IIb) above.
We have then $S_n^k=M_{n-k}^{alg}\,\max M_k
=M_{n-k}^{alg}M_{k}^{alg}\,=\,2^{\frac{n-2}{2}}$.
\end{itemize}
Thus, we have proved the Proposition for $n$ even.

To prove the Proposition for $n$ odd, we must calculate both
$S_n^k$ and $S_n^{nQM}$. We begin with $S_n^k$. Inserting
(\ref{mkpol2}) in the definition of $S_n$ for $n$ odd, we find \ba
S_n&=& \demi\,M_{n-k}\,M_k'\,+\,
\demi\,M'_{n-k}\,M_k\,.\label{snodd}\ea Without loss of
generality, we can suppose $k$ odd and $n-k$ even. Therefore, if
we assume correlations {\em k/(n-k)S}, $M_k$ and $M_k'$ can both
reach the algebraic limit due to property (IIb). So $S_n^k=\demi
M_k^{alg}\,\max(M_{n-k}+M'_{n-k})$; and due to property (IIa) this
maximum is $M_{n-k}^{alg}$. Thus $S_n^k=2^{\frac{n-3}{2}}$. Let's
calculate $S_n^{nQM}$. From the polynomial $S_n$ given by
(\ref{snodd}), we define the operator ${\cal{S}}_n$ in the usual
way. Therefore for the particular case $k=1$ we have \ba
\mbox{Tr}(\rho{\cal{S}}_n)&=&\demi\big[\mbox{Tr}(\rho\,{\cal{M}}_{n-1}\si_{a_n'})
+ \mbox{Tr}(\rho\,{\cal{M}}_{n-1}'\si_{a_n}) \big]\ea which is
bounded by $2^{\frac{n-2}{2}}$ because each of terms in the sum is
bounded by that quantity. This bound is reached by generalized GHZ
states, for suitable settings\cite{settings}. Therefore
$S_n^{nQM}=2^{\frac{n-2}{2}}$ for $n$ odd, and we have proved the
Proposition also for $n$ odd.\\

In conclusion: $n$-particle entanglement and $n$-particle
non-separability are {\em a priori} different concepts. We have
shown that it is possible to discriminate quantum entanglement,
not only against local variable models, but also against all
possible hybrid models {\em k/(n-k)S} allowing arbitrarily strong
correlations inside each subset but no correlation between
different subsets. Experiments aimed at demonstrating $n$-particle
entanglement should be analyzed using the generalized Svetlichny's
inequalities described in this work.\cite{mitchel}\\

This work was partially supported by the European Union project
EQUIP (IST-1999-11053) and by an ESF Short Scientific Visit Grant.

\begin{table}
\begin{center}
\vbox{
\begin{tabular}{c|c|c|c|c|c|}
   & $lv$ & {\em 2/1QM} & {\em 2/1S} & {\em 3QM} & {\em 3S} (alg.) \\ \hline
  $M_3$ & 1 & $\sqrt{2}$ & 2 & 2 & 2 \\
  $S_{3}$ & 1 & 1 & 1 & $\sqrt{2}$ & 2 \\
  prod. & 1 & $\sqrt{2}$ & 2 & $2\sqrt{2}$ & 4 \\
\end{tabular}
}
\caption{Maximal values of $M_3$ and $S_3$ under different
assumptions for the nature of the correlations (see text). The
last line is the product of the two values. }\label{tablefin}
\end{center}
\end{table}

\end{multicols}


\begin{thebibliography}{10}


\bibitem{machines} See for instance:
G. Alber {\em et al.},
  {\em Quantum Information: An Introduction to Basic Theoretical Concepts and
  Experiments}, Springer Tracts in Modern Physics {\bf 173} (Springer Verlag, Berlin, 2001).
\bibitem{mermin} N.D. Mermin, Phys. Rev. Lett. {\bf 65}
(1990) 1838
\bibitem{svet} G. Svetlichny, Phys. Rev. D {\bf 35}
(1987) 3066
\bibitem{mk} A.V. Belinskii, D.N. Klyshko, Phys. Usp.
{\bf{36}} (1993) 653; N. Gisin, H. Bechmann-Pasquinucci, Phys.
Lett. A {\bf 246} (1998) 1; R.F. Werner, M.M. Wolf, Phys. Rev. A
{\bf 61} (2000) 062102
\bibitem{note2}  See for instance: I. Percival, Phys. Lett. A {\bf
244} (1998) 495.
\bibitem{chshpaper} J.F. Clauser, M.A. Horne, A. Shimony, R.A.
Holt, Phys. Rev. Lett. {\bf 23} (1969) 880
\bibitem{werner} R.F. Werner and M.M. Wolf, PRA {\bf 64} (2001) 032112.
\bibitem{note1} Note that this implies that the bound for {\em lv}
is equal to the bound for {\em 1/1/.../1QM} (product states in
QM). From now onwards, we shan't distinguish these two cases
anymore.
\bibitem{cirel} B.S. Cirel'son, Lett. Math. Phys. {\bf 4} (1980) 83
\bibitem{sca1} V. Scarani, N. Gisin, J. Phys. A: Math. Gen.
{\bf 34} (2001) 6043
\bibitem{settings} 
To maximize $\frac{1}{2}\moy{{\cal{M}}_n+{\cal{M}}'_n}_{GHZ}$ for $n$ odd,
the $\si_{a_j}$ are taken of the form $\cos\alpha_j \si_x+ \sin\alpha_j
\si_y$. One possible choice for the settings is:
$\alpha_k=\alpha'_k+\frac{\pi}{2}$ for all $k$,
$\alpha'_1=...=\alpha'_{n-1}=0$,
$\alpha'_n=\frac{\pi}{4}$. For such settings, each correlation coefficient
becomes equal to $\frac{1}{\sqrt{2}}$ in modulus, with the good sign. See
\cite{mitchel}.


\bibitem{mitchel} P. Mitchel, S. Popescu and D. Roberts, in preparation.



\end{thebibliography}
\end{document}